\documentclass{aastex61}

\usepackage{graphicx}% Include figure files
\usepackage{dcolumn}% Align table columns on decimal point
\usepackage{bm}% bold math
\usepackage{rotating}
\usepackage{makecell}
\begin{document}

%\preprint{APS/123-QED}

\title{Night Vision for Small Telescopes}% Force line breaks with \\
\correspondingauthor{Robert Strausbaugh}
\email{robert.strausbaugh@asu.edu}
\author{Robert Strausbaugh}
\affil{%
Arizona State University, Department of Physics
}%
\author{Rebecca Jackson}
\affil{%
Arizona State University, School of Earth and Space Exploration
}%
\author{Nathaniel Butler}
\affil{%
Arizona State University, School of Earth and Space Exploration
}%
\date{\today}% It is always \today, today,

\begin{abstract}
We explore the feasibility of using current generation, off-the-shelf, indium gallium arsenide (InGaAs) near-infrared (NIR) detectors for astronomical observations.  Light-weight InGaAs cameras, developed for the night vision industry and operated at or near room temperature, enable cost-effective new paths for observing the NIR sky, particularly when paired with small telescopes.  We have tested an InGaAs camera in the laboratory and on the sky using 12 and 18-inch telescopes.  The camera is a small-format, 320x240 pixels of 40$\mu$m pitch, Short Wave Infra-Red (SWIR) device from Sensors Unlimited.  Although the device exhibits a room-temperature dark current of $5.7 \times 10^4$ $e^-s^{-1}$ per pixel, we find observations of bright sources and low-positional-resolution observations of faint sources remain feasible.  We can record unsaturated images of bright ($J=3.9$) sources due to the large pixel well-depth and resulting high dynamic range.  When mounted on an 18-inch telescope, the sensor is capable of achieving milli-magnitude precision for sources brighter than $J=8$.  Faint sources can be sky-background-limited with modest thermoelectric cooling.  We can detect faint sources ($J=16.4$ at $10\sigma$) in a one-minute exposure when mounted to an 18-inch telescope.  From laboratory testing, we characterize the noise properties, sensitivity, and stability of the camera in a variety of different operational modes and at different operating temperatures.  Through sky testing, we show that the (unfiltered) camera can enable precise and accurate photometry, operating like a filtered $J$-band detector, with small color corrections.  In the course of our sky testing, we successfully measured sub-percent flux variations in an exoplanet transit.  We have demonstrated an ability to detect transient sources in dense fields using image subtraction of existing reference catalogs.

\end{abstract}

\maketitle

%\tableofcontents

\section{Introduction}
The near-infrared (NIR), and particularly the short-wave-infrared (SWIR, $750-2500$ nm), are important wavelength bands in astronomy.  There are several advantages to working in these wavelengths. NIR light scatters off interstellar dust at a much lower rate than visible or ultraviolet light, because NIR wavelengths are much longer than the average size of interstellar dust particles \citep{IRastro}.  The redshifted NIR light that reaches us from cosmological distances probes the physics at shorter wavelengths in the rest frame unlike short wavelength light that would be absorbed by the intergalactic medium.  Also, many sources of astrophysical interest (e.g., low mass stars) are intrinsically quite red.

We are interested here in the potential uses of SWIR detectors on small telescopes to study transient astrophysical objects.
In the specific arena of time domain astronomy, there are several science cases for which NIR observations are advantageous.  Exoplanet transit light curves are likely to be more sharply defined in the NIR due to the effects of limb darkening \citep{limbtemp,limbwavelength}, making them easier to identify.  The study of exoplanets around smaller, cooler stars (M-type and Brown Dwarfs) is also optimal while operating in the NIR \citep{IRexo}.  Type Ia supernova light curves appear to be more standard in the NIR bands than at visible wavelengths \citep{SN_ir,sn_ir_reason}.  Gamma-ray Burst (GRB) afterglows have their peak brightness in the NIR, and they tend to fade more slowly in this regime \citep{afterglow,ag_length,ratir}.  Finally, the electromagnetic counterpart to gravitational waves from neutron star mergers should have a characteristic NIR signature \citep{kilo_ir}.

There are natural limitations for NIR observations. The terrestrial sky is very bright in the NIR, due to emission from particles, namely hydroxil (OH$^-$), in the atmosphere \citep{irsky_canada}.  As such, ground-based IR astronomy tends to be noise-limited by sky background.  This suggests an interesting instrument design path that utilizes inexpensive or off-the-shelf detectors, with higher-than-typical noise properties as compared to state-of-the-art detectors, because the detector noise can still be driven below the limiting sky noise in some situations.  We explore the implications for small telescopes, in particular, below.  Even when sky noise is not the limiting noise source, as in the case of very bright sources like exoplanet transits, detector stability and stability of the variable night sky in the NIR become key considerations.
  
  Astronomers have characterized well the NIR sky brightness, with expected magnitudes of $J=16.6$ per $arcsec^2$ and $H=15.5$ per $arcsec^2$ \citep{irsky_chile}.  It is possible to decrease the resulting sky background by utilizing narrow filters that sit in wavelength space between bright sky lines, which are also highly-time-variable.  The FIRE spectrograph on Magellan has achieved a mean inter-line sky continuum level of $Y=20.05\pm 0.04$, $J = 19.55\pm 0.03$, and $H = 18.80\pm 0.02$ (stat.) $\pm0.2$ (sys.) mag arcsec$^{−2}$ \citep{FIRE}.  A narrow $Y$-band filter could exploit one of these gaps (at 960-1080 nm), providing an uninterrupted window for observations \citep{yband}.  We note that all magnitudes presented in this paper are in the AB system.

\begin{figure}[h]
\includegraphics[scale=0.45]{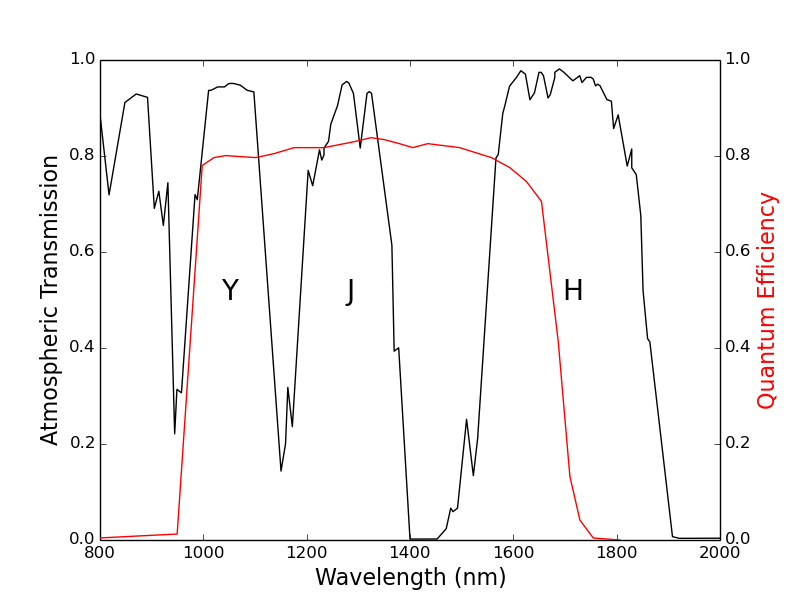}
\includegraphics[scale=0.37]{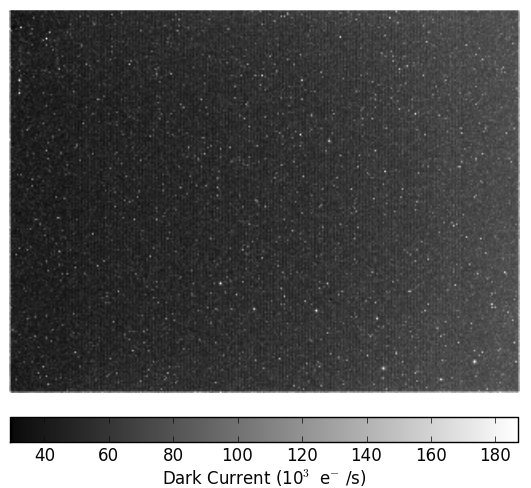}
    \caption{Left: An example InGaAs QE curve (from, http://www.sensorsinc.com), plotted in red over the atmospheric transmission spectra in black (from, http://modtran.spectral.com/modtran\_index), with the $Y$, $J$, and $H$ bands labeled.  Right: The visible gradient across a dark frame of the detector at 20 C, most likely due to a temperature gradient caused by the ohmic heating of the camera's electronics.}
\label{atmos_QE}

\end{figure}

The quantum efficiency (QE) of conventional silicon-based CCD detectors breaks down at wavelengths beyond 1000 nm \citep{otheringaas}.  The current standard for IR astronomy are HgCdTe detectors.  These detectors must be cryogenically cooled to decrease detector dark current to acceptable levels and to permit stable readout.  Astronomical instruments using these detectors tend to be both expensive and heavy.  A different semi-conductor, InGaAs (useful for 700-1700 nm) \citep{ingaas} \citep{ingaas2}, covers the shorter end of the NIR bandpass \citep[Figure \ref{atmos_QE};][]{HgCdTe}.  These detectors, which have become commercially available as a result of night vision industry, have decent noise properties at room temperature.   InGaAs detectors are cheaper to obtain than HgCdTe detectors, although the available format is currently smaller.  Depending on the application (and on the sky brightness per pixel in particular), there is the possibility of operating at relatively high temperature, at or near room temperature.

In this study, we characterize a commercially available SWIR camera from the Goodrich corporation in both laboratory and on-sky settings with small telescopes and realistic observing conditions.  Below, we show the results of the laboratory testing, including dark rate (and its behavior with temperature), gain, read noise, QE, linearity, and charge persistence.  We also present the results of testing the InGaAs detector on the sky. We show the detector's photometric precision, its color (comparing $J$ band to $Y$ and $H$ bands), and finally present a light curve of HD189733, which shows the predicted dip caused by a known exoplanet transit.  Having presented these results, we show that we can account for the noise present in our system by accurately modeling the statistical sources of noise. Using our model for noise, we put limits on what sources we can study.  

\section{Camera Description and Laboratory Testing}
\label{sec:lab_test}

We have tested a small-format (320x240 pixel) Short Wave Infra-Red (SWIR) camera from Sensors Unlimited, Inc.\footnote{http://www.sensorsinc.com}, a division of UTC Aerospace Systems.
The SU320HX-1.7RT is a Mil-Rugged InGaAs video camera featuring high-sensitivity and wide operating temperature range.
It has a compact size ($<$ 3.8 in$^3$) and can be operated over a wide temperature range (-40 to 70 C) with low required power ($<$ 2.9 W at 20 C).
The sensor has large pixels (40$\mu$m x 40$\mu$m) and is advertised to have high pixel operability ($>99$\%) and high sensitivity ($>65$\% QE) from 900 nm to 1.7 $\mu$m.  The full-well depth is $10^7$ e$^-$.
A built-in thermo-electric cooler (TEC) is designed to maintain a stable sensor temperature of 20 C.

The analog signal from the sensor is digitized to 12-bit data in CameraLink format.
We use a frame-grabber from National Instruments (NI PCIe-1427) and the NI-IMAQ software\footnote{see, http://www.ni.com} for image acquisition.  Custom python scripts have been written to provide a GUI and scripting interface as well as to provide real-time image visualization; this software is called ICACTI (Infrared Camera for Astrophysical and Cosmological Transients Interface) and and has been made freely available\footnote{https://github.com/rstrausb/ICACTI}.  We set camera modes using the serial interface and use the NI-IMAQ C libraries to store image frames in FITS format.  
We operate the camera in a continuous read mode of individual 16.3 ms frames which are summed into longer exposure frames as desired.  The 16.3 ms frame time is found to offer an acceptable compromise: the noise floor is well-sampled while there is also sufficient dynamic range to avoid saturation due to bright stars.  In the sub-sections below, we discuss laboratory measurements of the detector dark current, gain, quantum efficiency, persistence, and linearity.

\subsection{Dark Current, Read Noise, and Sensor Gain}
Blocking light to the camera, we measure a dark current at the nominal operating temperature (20 C) of $5.7\times10^4$ $e^-s^{-1}$ per pixel (i.e., $3.6\times10^{13} e^-s^{-1}m^{-2}$).  A variation in the dark current level is found to be present across the sensor, similar to the pattern discussed for the device in \citet{otheringaas}.  We find that this gradient persists when the internal TEC is turned off and cannot, therefore, be due to the TEC.  A likely explanation for the temperature gradient is ohmic heating due to circuitry behind the sensor.

Similar detectors have been characterized \citep{otheringaas,taiwaningaas}, with similar dark rates at 20 C.  The camera studied by \citet{otheringaas} has a dark rate of $3.5\times10^{13} e^-s^{-1}m^{-2}$. \citet{taiwaningaas} tested an InGaAs detector with a dark rate of $6.6\times10^{14} e^-s^{-1}m^{-2}$.

\begin{figure}[h]
\includegraphics[scale=0.45]{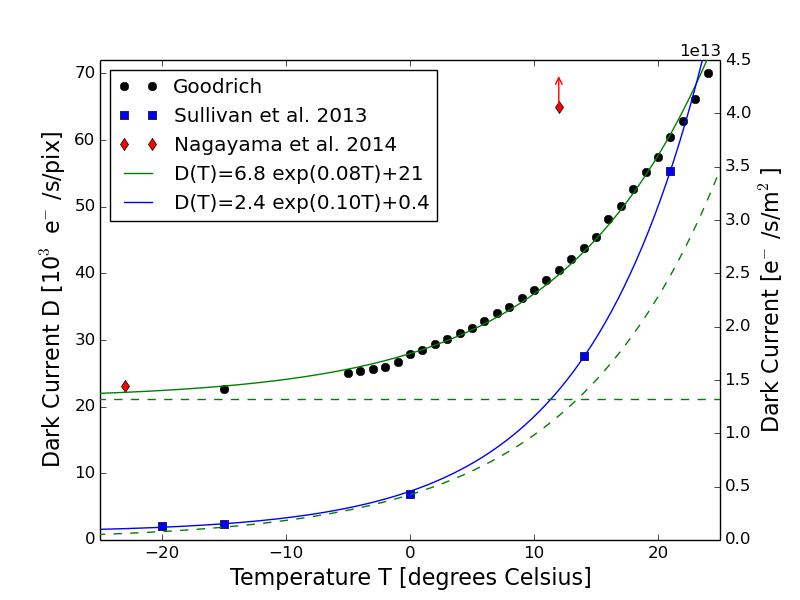}
\includegraphics[scale=0.45]{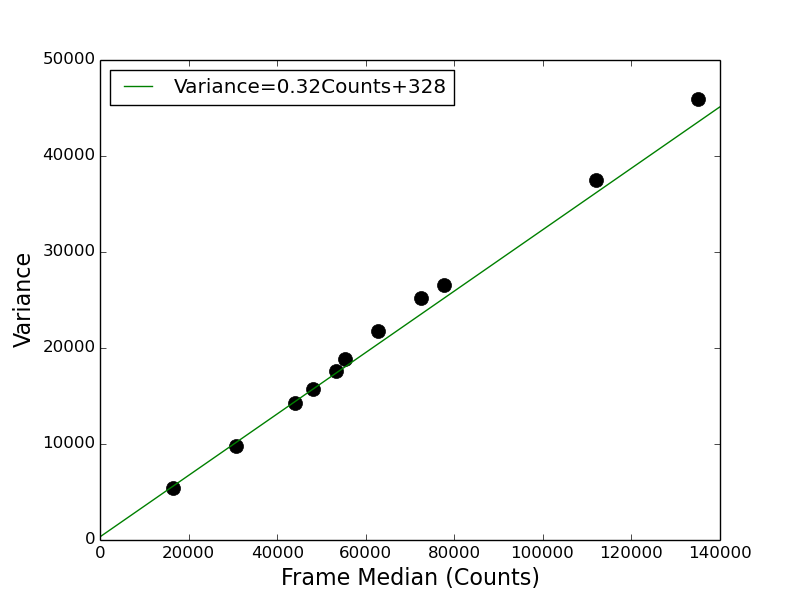}
    \caption{\textit{Left:} The effect of temperature on dark rate, utilizing an external 3-stage TEC.  The units on the right side y-axis denote the per pixel dark rate if each detector had the same sized pixels (40 $\mu$m $\times$ 40 $\mu$m). The left y-axis shows the dark rate in terms of area instead of pixels, so that the different detectors can be directly compared.  \textit{Right:} The relationship between variance and signal is plotted, with the equation describing the fit shown.}
\label{cooling}

\end{figure}

We have also experimented with additional cooling (Figure \ref{cooling}) using an external TEC mounted to the side of the aluminum camera housing.
We employed a Ferrotec 3-stage Deep Cooling unit, capable of generating a $\Delta T = 111$ C temperature differential between the hot and cold side of the TEC.  The entire camera was kept near 0 C in an external, cooled enclosure and insulation was wrapped around the device and 3-stage TEC.  However, we did not achieve the expected 111 C temperature differential due to the lack of direct contact between the 3-stage TEC and sensor.
By measuring both the signal level (dominated by dark current; Figure \ref{cooling}R) and the signal variance in several frames captured over a range of temperatures (Figure \ref{cooling}L),
we find an inverse gain of approximately 3 e$^-$/ADU.  We find a read noise of about 12 e$^-$ per 16.3 ms frame.

%Discuss the dark current rate at another interesting temp (e.g., 0 C) and compare what you find there to what is found in the other papers.  Are we still lower?  Then also project down to a very low temp which drives the dark below the J-band sky for some nominal pixel plate scale and reference the discussion below.

After moderately cooling the camera (down to -15 C), we achieve a dark current of $1.42\times 10^{13} e^-s^{-1}m^{-2}$.  \citet{otheringaas} and \citet{taiwaningaas} achieve a more significant decrease in the dark current after similar cooling ($1.5 \times 10^{12} e^-s^{-1}m^{-2}$ and $1.4\times10^{13} e^-s^{-1}m^{-2}$, respectively).
We have modeled the change in dark current with respect to temperature using simple exponential functions, of the form $D(T)=A e^{B T}+C$, where $D(T)$ is the dark current as a function of temperature, $T$.  These exponentials are plotted as a solid green line, for the Goodrich detector, and as a solid blue line, for the \citet{otheringaas}, in Figure \ref{cooling}.

Separating the exponential and constant baseline (the dotted green components in Figure \ref{cooling}), demonstrates that both our Goodrich detector, and the \citet{otheringaas} detector have similar exponential behavior, with a large discrepancy between the constant offsets.
The large constant component still present when cooling the Goodrich detector is most likely due to the fact that we did not directly cool sensor, and instead cooled the entire camera unit.  

Directly cooling the sensor inside the Goodrich camera would require to disassembly of the camera.  We would need to run longer wires from the electronics to the sensor, attach the TEC to the back of the sensor, and run water cooling to pull heat from the hot side of the TEC.  This would all need to be enclosed in a larger aluminum case, with fans to dump heat from the water cooling system.   Although not yet developed, such a scheme to apply direct cooling to the sensor would likely remove the pattern noise found in the dark frames.  An external triggering device could also be housed in the new camera assembly (as motivated in Section \ref{disc}).

Cooling the sensor directly, we expect to achieve a dark current on the order of the sky background ($\approx 6000 e^-s^-$ per pixel in J-band for the telescopes we have utilized; see Section \ref{disc}).  We can determine at which temperature this dark current will occur using our exponential fit models; however, this temperature is very sensitive to baseline level.  If we assume a similar baseline to the \citet{otheringaas} detector, we should achieve a dark current comparable to the sky background level at $T=0C$. Even at a baseline level several times higher than the \citet{otheringaas} detector, the temperature needed to achieve sky background levels in the dark current would still be well within the cooling range of the Ferrotec TEC.

We note that our inferred inverse gain value is somewhat smaller than the manufacturer's value quoted for 16.3 ms frames.  This suggests some smoothing present in the analog-to-digital conversion.
By calculating the autocorrelation between subsequent frames on a pixel-by-pixel bases, we determine that approximately 15 subsequent frames show signs of correlation.  We expect that this smoothing is due to the capacitors present in each pixel read-out.  Our measurements of all astronomical sources below were conducted using 1 s integration sums of the 16.3 ms frames, sufficiently long enough to average over this capacitive smoothing.  
We estimate that the effective inverse gain in a 16.3 ms frame is approximately 5 e$^-$/ADU.

Using Figure \ref{cooling}, we are also able to determine the read noise of the detector.  Using the y-intercept, and converting to appropriate units, the read noise corresponds to 12 $e^-frame^{-1}$ (90 $e^-s^{-1}$).

Finally, we note that cooling the InGaAs detector strongly comes at a price: the sensitivity to longer wavelength light is degraded for InGaAs (e.g. Figure 4-8\footnote{https://www.hamamatsu.com/resources/pdf/ssd/infrared\_kird9001e.pdf}.  The sensitivity lost by this detector would occur in the $H$-band, where the sky brightness level is high, perhaps making the loss of sensitivity at longer wavelengths acceptable.

\subsection{Quantum Efficiency, Persistence, and Linearity}

We confirmed the advertised QE of the camera by measuring monochromatic light with both the Goodrich camera and a photo-electric diode, with known responsivity.  We measure a QE of $>80\%$ between 950-1050 nm and $>60\%$ between 1050-1700 nm.  This wavelength range encompasses the entire $J$-band and most of the $H$-band with moderate efficiency ($>60\%$), and importantly, the very clean $Y$-band at high efficiency ($>80\%$).

The effects of charge persistence can be important for time domain astronomy, in particular.  To quantify the persistence, while the detector was collecting data, a light source was turned on and off.  We fit exponentials to this data, resulting in a time constant of 23.9 ms for exponential growth (when the light was turned on) and 16.5 ms for decay (when the light was turned off).  These time scales are on the order of the individual frame time (16.3 ms) and are likely to be much shorter than any natural timescales for typical astrophysical transients.

We also tested the linearity of our detector to ensure it was suitable to study the wide range of magnitudes inherent in transient astronomy: from bright stars hosting exoplanetary systems, to dim and distant SNe and GRBs.  The Goodrich detector exhibits a linear response over a range of 10 e$^-$ per pixel to $3 \times 10^6$ e$^-$ per pixel (a dynamic range of $>10^5$).

\section{Sky Testing}

In order to verify our laboratory device characterization just discussed, we conducted a number of on-sky imaging campaigns (Table \ref{tabl}).  In addition to confirming device properties utilizing a noise model for sources detected by the camera, which we explore in Section \ref{sec:noise} below, there were two major goals of these campaigns: (1) to conduct proof of principle observations of both very bright sources and faint sources near the noise floor of the device, and (2) to observe a sufficient number of field stars to allow for the photometric characterization of the camera in terms of flux and color accuracy.

We mounted the Goodrich camera on an 18-inch ($f/4.5$) and 12-inch ($f/10$) telescope and conducted sky testing from a roof top on ASU's campus in Tempe, Arizona.  The sensor has a plate scale of 4.0 arcseconds/pixel on the sky on the 18-inch telescope, with a field of view of 21.4 x 16.0 arcmin$2$.  The sensor has a plate scale of 2.7 arcseconds/pixel on the 12-inch telescope, with a field of view of 14.4 x 10.8 arcmin$2$.  Due to the large size of the camera's pixels on the sky, it is sometimes true that the entire full width at half maximum (FWHM) of a star is contained in a single pixel.  At site with better seeing, this is likely to be a common occurrence.   Single pixel source monitoring could be useful for some high-precision photometric applications which seek to mitigate the effect of intra-pixel and pixel-to-pixel gain variations.

The 18-inch Newtonian telescope, manufactured by JMI, features a highly stable, 36 inch split ring polar mount.  The primary mirror and secondary diagonals used for telescope are supplied by Galaxy Optics (Buena Vista, CO), which produce high quality, large diameter Newtonian mirrors. It is a precision annealed, 2 inch thick pyrex primary mirror floated on 18-points, which provides even support and prevents pressure areas leading to distortion. The mirrors are manufactured to yield RMS wavefront errors below that of the diffraction limit. The optical coating are custom fit to be effective in the IR band, with a $<$ 1/100 wave center to thickness variation and mirror reflectivity of 98$\%$.  The 12-inch telescope is an LX-200 Cassegrain telescope from Meade.

Data for the sky tests was collected and saved from the camera using the acquisition software (Section \ref{sec:lab_test}).  The data were then analyzed using a pipeline similar to the one used for RATIR \citep[Reionization And Transients Infrared/Optical camera;][]{RATIR_pipeline}, as follows.  Images were first reduced using flat-fielding algorithms in Python.  Stars in the reduced images are found using Source Extractor \citep{sextractor}, and images are aligned based on those star locations using astrometry.net \citep{astrometry}.  These aligned images are then stacked using SWARP \citep{swarp}.  Finally, photometry is obtained using Source Extractor on the stacked images.

\begin{figure}[h]
\includegraphics[scale=0.35]{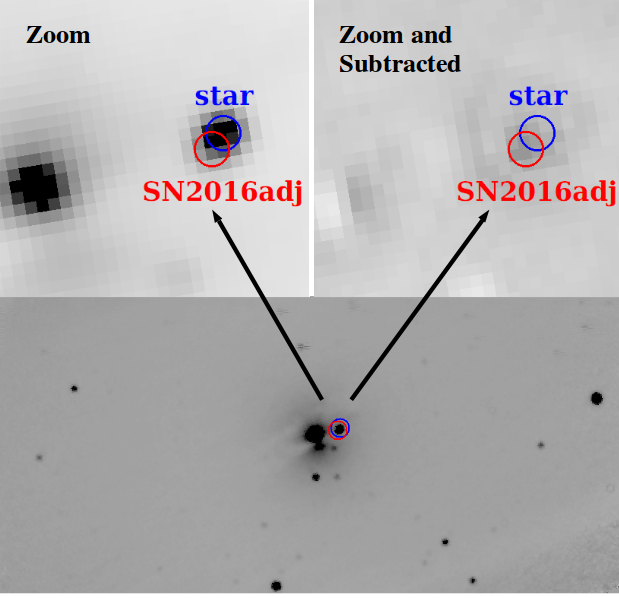}
\includegraphics[scale=0.6]{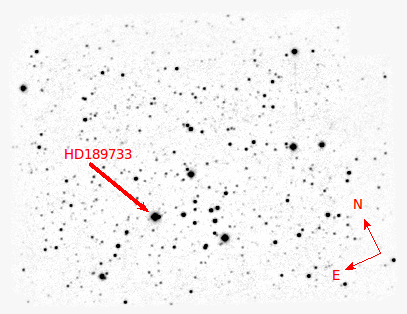}
\caption{\textit{Left:} The field of Centaurus A.  Centaurus A is the host galaxy for the recent supernova, SN2016adj. The images at the top are a zoom on the region of SN2016adj, and an image subtraction of the same zoomed area of the sky. The image subtraction removes the foreground star, revealing SN2016adj.  \textit{Right:} The field containing the exoplanet orbiting HD189733.  The HD189733 system has a transiting exoplanet.  The light curve showing a detection of the exoplanet is shown in Figure \ref{exo_lightcurve}.  Additional information for these images can be found in Table \ref{tabl}.}
\label{fields}
\end{figure}

The results of running our pipeline on data collected for the fields of the galaxy Centaurus A and HD189733 are shown in Figure \ref{fields}.  Additional information about the data from these fields, as well as the field HAT-P-36, are shown in Table \ref{tabl}.

\begin{table}[h!]
    \caption{Summary of Astronomical Fields Observed for On-Sky Testing}
\begin{tabular}{ c c c c c c c}
 Figure Label & Target & \hspace{-13 mm} \makecell{RA (center) \\ Dec (center)} & \hspace{-10 mm} \makecell{Start Time (UTC) \\ End Time (UTC)} & Effective Exposure Time (s) & Telescope \\
\hline
 2013\_10\_29 & HD 189733 & \hspace{-13 mm} \makecell{$20^h00^m30.189^s$ \\ $+22^{\circ}43^{'}26.867^{"}$} & \hspace{-10 mm} \makecell{2013-10-30T03:24 \\ 2013-10-30T04:55} &  463 & 12-inch Meade  \\
  2016\_02\_16 & Centaurus A & \hspace{-13 mm} \makecell{$13^h25^m37.200^s$ \\ $-42^{\circ}59^{'}44.160^{"}$} & \hspace{-10 mm} \makecell{2016-02-16T10:28 \\ 2016-02-16T13:20} & 8198 & 12-inch Meade  \\  
 2016\_04\_14 & HAT-P-36 & \hspace{-13 mm} \makecell{$12^h33^m35.204^s$ \\ $+44^{\circ}55^{'}00.796^{"}$} & \hspace{-10 mm} \makecell{2016-04-15T05:21 \\ 2016-04-15T09:46} & 14610 & 18-inch JMI \\
 2016\_04\_26 & HAT-P-36 & \hspace{-13 mm} \makecell{$12^h33^m15.343^s$ \\ $+44^{\circ}53^{'}26.867^{"}$} & \hspace{-10 mm} \makecell{2016-04-27T04:41 \\ 2016-04-27T08:33} & 12245 & 18-inch JMI\\ 
\hline
\end{tabular}
\label{tabl}
%\\ \textbf{ Table 1. } \text{Additional information for the data plotted in Figures \ref{fields}, \ref{photometric_performance}, and \ref{col_corr}.  }
\end{table}

We targeted HD189733 and HAT-P-36 as they are known to host exoplanets.  Centaurus A was imaged a week after the detection of SN2016adj \citep{sn2016adj}.  Despite blending with a nearby ($J=10.8$ mag) star, image subtraction with HOTPANTS \citep{HOTPANTS}, using a convolved 2MASS $J$-band archival image as a reference, reveals the $J\sim 13$ mag supernova (Figure \ref{fields}).

We collected the HD189733 data on the 12-inch telescope before the 18-inch telescope was operational.  The Centaurus A data were collected on the 12-inch telescope due to the galaxy's location on the sky and the fact that it would have been challenging to point our 18-inch telescope that far South.

In the following sub-sections, we use the data from HD189733, Centaurus A, and HAT-P-36 fields to determine the Goodrich camera's performance on the sky, testing its photometric performance, comparing its color to the established 2MASS \citep{2mass} and Pan-STARRS \citep{panstarrs} catalogs, and determining a color correction term to compare our broadband results to these filtered catalogs.  We are able to detect exoplanet transits, as evidenced by the dip in the lightcurve of HD189733 (Figure \ref{exo_lightcurve}), associated with the transit of exoplanet HD189733b.

\subsection{Photometric Performance}

%Multiple fields have been imaged in preparation for future observations of exoplanet transits, namely the fields of WASP-33 and HAT-P-36.  The photometric performance of the InGaAs detector on stars in both fields is shown in Figure \ref{photometric_performance}.

\begin{figure}[h!]
\includegraphics[scale=0.6]{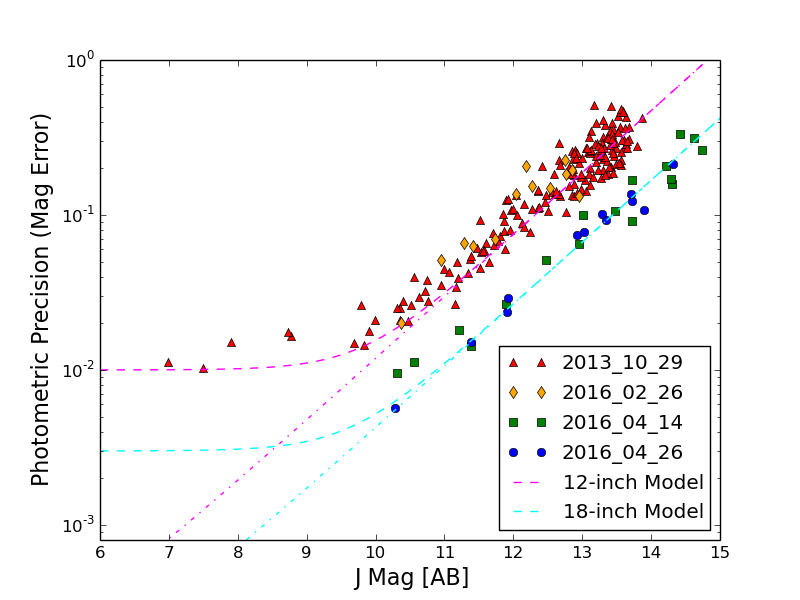}
\caption{The photometric errors of detected stars compared to a model of statistical noise, described in Section \ref{disc}; the dashed line shows a model with a systematic error term, and the dot-dashed line shows the same model without the systematic term.  The observations on these nights are detailed in Table \ref{tabl}.}
\label{photometric_performance}
\end{figure}

We obtained photometry from as many stars from the HD189733, Centaurus A, and HAT-P-36 fields as possible.  In Figure \ref{photometric_performance}, we have plotted the apparent magnitude of the fields stars against their respective errors.

The theoretical curves plotted in Figure \ref{photometric_performance}, modeling the 18-inch and 12-inch telescopes, are calculated in Section \ref{disc}.  We note that the models closely match the data, and as such we can use these models to predict whether an exoplanet transit will be visible for stars of a certain magnitude.

Following the dot-dashed curve for the 18-inch telescope in Figure \ref{photometric_performance} to brighter sources, we find that milli-magnitude precision should be possible for stars brighter than $J=8$.

\subsection{Color Correction}

If the camera is used without a bandpass filter, as it was in the sky testing presented in this paper, a color-correction term may be needed to compare the measured magnitude of sources to the $J$, $H$, and $Y$ band measurements from established catalogs.  We compare our data from the nights described in Table \ref{tabl} to catalogs from 2MASS for $J$ and $H$ bands, and Pan-STARRS for $Y$ band.  There are no Pan-STARRS data, however, for the Centaurus A field, as that survey did not collect data south of declination -30 degrees.

\begin{figure}[h!]
\includegraphics[scale=0.45]{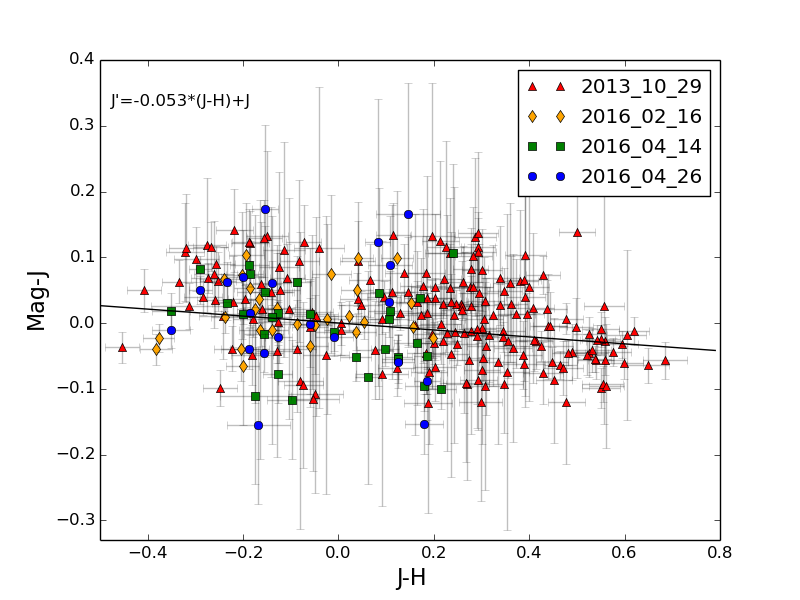}
\includegraphics[scale=0.45]{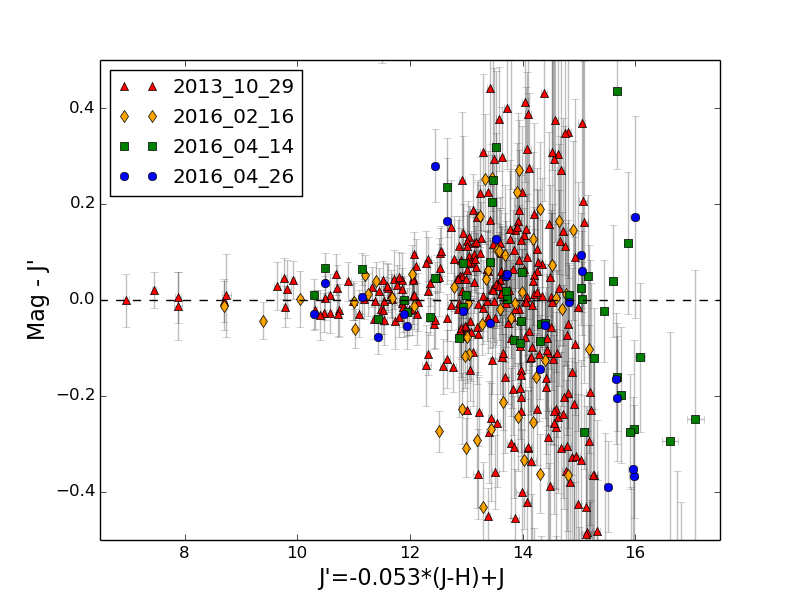}
\caption{\textit{Left:} A color correction term is given by the slope of the line (-0.05) through the data.  This term is important for comparing our instrumental magnitudes with catalog magnitudes ($J$ and $H$ taken from the 2MASS catalog).  $J^\prime$ is the expected magnitude, given $J$ and $H$ from the catalog and applying the color correction term.  \textit{Right:} The accuracy of the color correction is shown by comparing our measured magnitude relative to the expected magnitude $J^\prime$.  The observations on these nights are detailed in Table \ref{tabl}.}
\label{col_corr}
\end{figure}

Comparing our magnitude to the $J$ and $H$ bands from 2MASS yields a color term of -0.05$\pm$ 0.03 (Figure \ref{col_corr}).  This small color correction term demonstrates that the bulk of light collected by the camera is in the $J$-band, with a small fraction of light in the $H$-band.

The overall photometric accuracy is plotted in the right-most graph in Figure \ref{col_corr}.  Including the color term derived above, the photometry is accurate (within the error bars) for both bright ($J\leq 12$) stars and fainter stars where the uncertainties are large.

\begin{figure}[h!]
\includegraphics[scale=0.45]{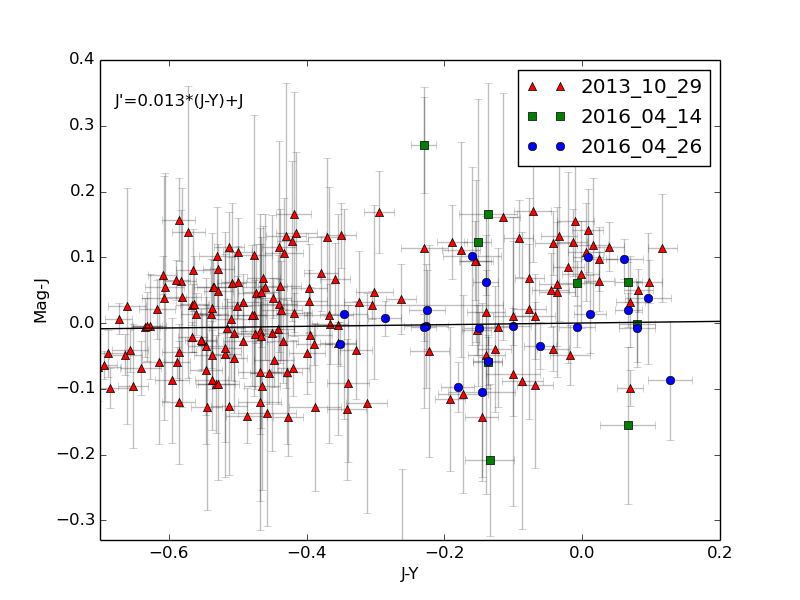}
\includegraphics[scale=0.45]{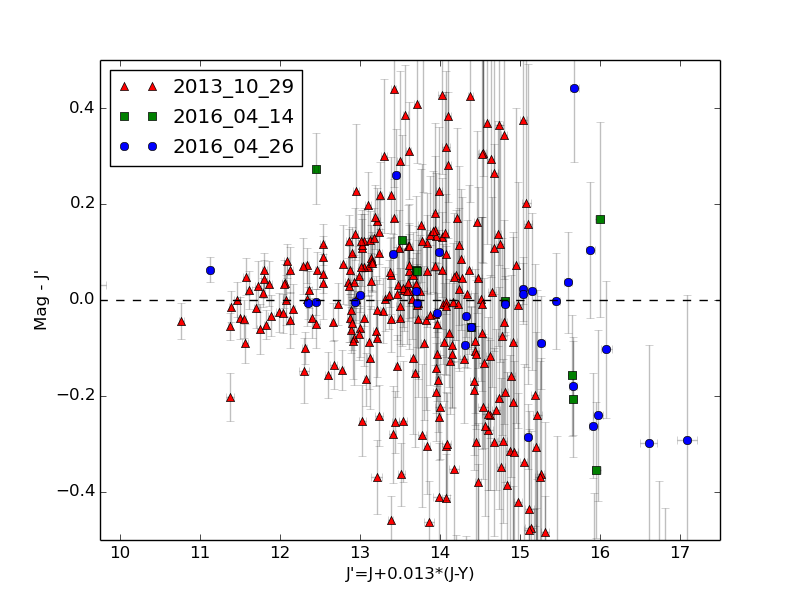}
\caption{\textit{Left:} The color correction term is given by the slope of the line (0.01) through the data.  This term is important for comparing our instrumental magnitudes with catalog magnitudes ($J$ taken from the 2MASS catalog and $Y$ taken from Pan-STARRS).  $J^\prime$ is the expected magnitude, given $J$ and $Y$ from the catalogs and applying the color correction term.  \textit{Right:} The accuracy of the color correction is shown by comparing our measured magnitude relative to the expected magnitude $J^\prime$.  The observations on these nights are detailed in Table \ref{tabl}.}
\label{colorY}
\end{figure}

There is also potentially a color correcting in the blue due to the detector response shortward of $J$-band (Figure \ref{atmos_QE}).  Comparing our magnitudes with the $J$-band from 2MASS and the $Y$-band from Pan-STARRS, we derive a color correction term of 0.01$\pm$ 0.03 (consistent with zero), as seen in the left graph of Figure \ref{colorY}.

With the small $J-H$ color correction and the even smaller $J-Y$ color correction of (-0.05 and 0.01, Figures \ref{col_corr} and \ref{colorY}, respectively), we note that our (unfiltered) camera acts very much like a filtered $J$-band camera.  There appears to be an overall blue-ward shift in the color data when using the 12-inch telescope compared to the 18-inch telescope (or a red-ward shift in the 18-inch telescope compared to the 12-inch telescope) as seen in Figures \ref{col_corr} and \ref{colorY}.

\subsection{HD189733b Transit}

The exoplanet HD189733b was detected around its host star by \citet{hd189733b} using radial velocity measurements.  It was verified spectroscopically shortly thereafter by the same group \citep{hd189733b}.  HD189733 is a well studied system, due to the brightness of the star \citep[$J$ and $H\approx7$;][]{2mass} and the depth of the transiting exoplanet \citep[a $\gtrsim$2\% decrease in star brightness;][]{hd189733b}.

\begin{figure}[h!]
\includegraphics[scale=0.45]{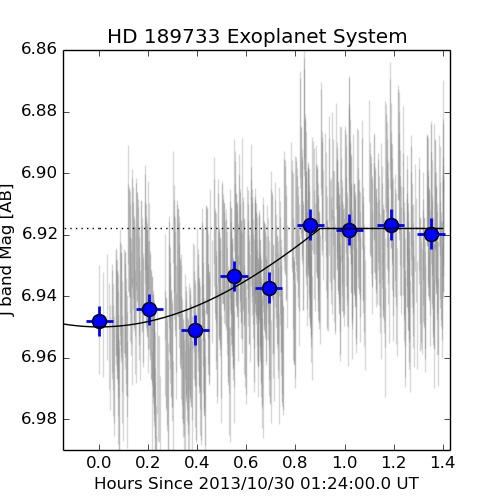}
\caption{Unfiltered observation of the transit of the 7th magnitude system HD 189733.  This proof of concept observation was taken with the Goodrich detector on a rooftop of a building on the Arizona State campus in Tempe, AZ.  We identify the correct transit depth and end time.  The data were taken with a 12 inch Meade telescope with a plate scale of 2 arcsec/pixel.}
\label{exo_lightcurve}
\end{figure}

Despite non-ideal conditions (bright sky, hot buildings, etc.), the latter half of one transit was recorded.  Our light curve is shown in Figure \ref{exo_lightcurve}.  Additional information about this observation can be found in Table \ref{tabl}.  Having demonstrated that we are capable of detecting exoplanet transits, we have continued to work with the Goodrich detector to obtain exoplanet light curves.  Lightcurves for the HAT-P-33 system, as well as several others, will be presented, in conjunction with data taken simultaneously with RATIR, in a follow-up paper.

\section{Discussion of Noise Properties} \label{disc}
\label{sec:noise}

In order to characterize the quality of our data acquisition and to allow for future observation planning, we must understand the sources of noise for our detector.  Equation \ref{s/n} summarizes the expected signal-to-noise ratio ($S/N$) as a function of sources of signal in the detector:

\begin{equation}
\frac{S}{N}=\frac{N_*}{\sqrt{N_* + n_p N_D + n_p N_S + n_p D_T + n_p RN^2 + (\sigma N_*)^2}},
\label{s/n}
\end{equation}
where $N_*$ is the electron count from the source during the exposure time, $N_D$ is the number of dark current electrons from the sensor, $N_S$ is the sky brightness in electrons, $N_T$ is the number of electrons due to the thermal emissions of the instrument (telescope, camera window, etc.), $RN$ is the read noise, and $\sigma$ is a systematic term to represent any errors that scale with the source brightness.  The factor $n_p$ is the number of pixels used to extract a source from an image.  Here, we are assuming the counts ($N_D$, $N_S$, $N_*$, and $N_T$) are Poisson distributed.  The read noise is squared in the noise calculation as it is a Gaussian noise source; the systematic term, with error proportional to $\sigma$, is also assumed to be Gaussian.  Having thoroughly tested the camera in the laboratory and on the sky, we can now compare our observed measurement uncertainties to calculations of the noise properties, using Equation \ref{s/n}.

Assuming we cool the camera with its internal TEC ($T\approx$15 C), the dark current, $N_D$, is $5.2\times10^4$ $e^-s^{-1}pixel^{-1}$.  Cooling the camera with an external TEC to $-5C$ reduces the dark rate to $2.5\times10^4$ $e^-s^{-1}pixel^{-1}$, as seen in Figure \ref{cooling}.  It should be noted that we believe this is not the actual temperature of the sensor, but instead the temperature inside the camera; if we were able to directly cool the sensor, the dark should be much lower at $-5C$.

We note that the color plots in Figures \ref{col_corr} and \ref{colorY}, suggest that our detector operates almost exclusively in the $J$-band.  Assuming the data from Las Campanas in Chile as a best case scenario, the sky background, $N_S$, has a level of 6600 $e^-s^{-1}pixel^{-1}$ in the $J$-band \citep{irsky_chile}.

In Figure \ref{cooling}, we show that the read noise ($RN$ in Equation \ref{s/n}) of the Goodrich detector is 12 $e^-pixel^{-1}frame^{-1}$.  We currently capture data with a frame time of 16 ms, which means that 60 of these frames are added to create a one second exposure.  Adding 60 frames together brings the read noise up to 90 $e^-pixel^{-1}s^{-1}$.  This read noise is on the order of the sky background, but is much smaller than the contribution of dark noise.  The level of the read noise can be reduced dramatically by externally triggering exposures for the entire one second.  We have verified this in the laboratory but do not typically use external triggering in our on-sky setup.

Assuming the equipment in the experimental setup acts as a black body, the thermal noise registered by the sensor would be 4850 $e^-s^{-1}$ across the entire collecting area.  On a pixel level, this noise, $N_T$, is negligible ($<1e^-s^{-1}pixel^{-1}$).

The remaining sources of noise in Equation \ref{s/n} depend on $N_*$, the flux from the source itself.  For a source of a given magnitude, the flux can be calculated using Equation \ref{mag},

\begin{equation}
mag_{AB}=-2.5\log({\frac{F}{3631\times10^{-23}}})
\label{mag}
\end{equation}
where $F$ is the flux from the source, and $3631\times10^{-23}$ is a conversion factor from Jansky to cgi units.  This flux can be converted into a signal on the detector using Equation \ref{flux},

\begin{equation}
F=\frac{h\times \nu \times g}{d\nu \times A \times {QE}}\times\frac{C}{dt}
\label{flux}
\end{equation}
where $h$ is Plank's constant, $\nu$ is the frequency of light, $g$ is the inverse gain of the camera, $d\nu$ is the bandwidth over which the observation is done, $A$ is the collection area of the telescope, $QE$ is the quantum efficiency of the camera and telescope together, $C$ is the number of counts on the detector, and $dt$ is the integration time over which the data are collected.  

The inverse gain of the camera is 5 e$^-$ per count (Section \ref{sec:lab_test}).  The quantity $\nu/d\nu$ is about $0.24$ in the $J$-band, which was the dominant color as seen in Figures \ref{col_corr} and \ref{colorY}.  When the mirrors of the telescope, both primary and secondary, are taken into account, a conservative estimation of the total $QE$ is around 20\%.  The two telescopes used for testing have an 18-inch and 12-inch primary mirror respectively, which is used to calculate the area, A.  The quantity $C/dt$ is either the count rate from a source of interest or is taken as the number of counts per second from the statistical sources of noise to determine limiting magnitudes.  

Given the magnitude of a source, we can calculate the theoretical contribution to the noise by statistical sources, using the denominator of Equation \ref{s/n}.  This is shown in Figure \ref{photometric_performance}, with the ``Model'' curves for both the 18-inch telescope and the 12-inch telescope.  The dashed line models in Figure \ref{photometric_performance} include the systematic term, $\sigma$, from Equation \ref{s/n}, while the dot-dashed line models do not include the systematic term.  The data from the 12-inch telescope in Figure \ref{photometric_performance} (the red triangles and orange diamonds) show that there is systematic uncertainty preventing precision better than 10 mmag.  Our sensitivity is somewhat better with the 18-inch telescope.  In any case, we are confident that this systematic term is not due to the camera, but instead due to poor observing conditions (very bright sky in the Phoenix, AZ metro area).  Evidence for this can be seen in Figure \ref{cooling} (e.g., the right-most point in the right panel), where we demonstrate stability in the laboratory to better than 1.5 mmag.

The dot-dashed lines have been included in Figure \ref{photometric_performance} to show the precision we can expect at a darker site.  The dashed line theoretical curves go through the observational data points at lower magnitudes (due to the systematic term, $\sigma$).

Optimal ground-based observations tend to be sky-noise-limited.   In order for our detector to operate in a regime dominated by sky noise, the detector would need to be cooled directly, which would involve redesigning the camera housing.  Following the trend seen in Figure \ref{cooling}, the Goodrich detector would need to be cooled to 0 C in order to have a dark signal approximately equal to sky emission.   As mentioned above (Section \ref{sec:lab_test}), cooling the sensor significantly may lead to a loss of sensitivity at longer wavelengths.  This trade-off could be acceptable due to the small fraction of $H$-band light detected, as seen in Figure \ref{col_corr}, and the higher sky noise in the $H$-band.

\section{Conclusion}

We have thoroughly tested a Goodrich InGaAs detector in the laboratory and on the sky for use in transient astronomy.  Our laboratory testing (e.g., Figure \ref{cooling}), indicates that the Goodrich detector performs similarly to previously tested InGaAs detectors.  At room temperature (20 C), the Goodrich detector has a dark rate of 57,000 $e^-s^{-1}$ per pixel.  We determined the read noise of the detector to be $12 e^-$ per frame, and the gain of the detector to be 5 $e^-$ per count (Figure \ref{cooling}).  The QE (Figure \ref{atmos_QE}) was confirmed to be between 60-90\% over a wavelength range that includes $Y$, $J$, and parts of $H$ band (900-1700 nm).  Due to large pixels and highly-stable readout, the detector's response is linear over a factor of $>10^5$ in dynamic range.

Through sky testing, we conclude that the unfiltered detector yields photometry comparable to a filtered $J$-band detector.  Comparing our data to 2MASS and Pan-STARRS, we derive a $J-H$ color term of -0.05$\pm$ 0.03 (Figure \ref{col_corr}) and a $J-Y$ color term of 0.01$\pm$ 0.03 (consistent with zero; Figure \ref{colorY}).  We have shown that we are able to successfully model the noise present in our system (Figure \ref{photometric_performance}), and using that model we can predict whether or not we will be able to study certain transient sources.  By catching the tail-end of the transit of HD189733 (Figure \ref{exo_lightcurve}), we have shown that the Goodrich camera is capable of detecting exoplanet transits and that our data-reduction pipeline is capable of extracting meaningful light curves, with better than 1\% photometry, from the data.  Even though under-sampled, the images are amenable to image subtraction using existing 2MASS catalog data, making possible faint source identification in potentially crowded fields (i.e., SN2016adj in Figure \ref{fields}).

According to the noise model fits to our data (Figure \ref{photometric_performance}), we expect to achieve milli-magnitude precision for $J<8$ sources on an 18-inch telescope.  This level of precision is achieved without any advanced dithering routines, such as the snapshot technique \citep{dither}, or any additional cooling.  Implementing these would potentially push our noise ceiling down to a regime dominated by sky background.  Overall, we find that mounting to smaller telescopes has the benefit of allowing for a larger area of the sky to be imaged, while also allowing for more sky background to potentially dominate the dark noise at each pixel.  With the InGaAs camera mounted to a larger telescope, a finer resolution on the sky is possible; however this combination will tend to lead to a noise budget dominated by dark noise.

Based on our work with the InGaAs detector in the laboratory and on the sky, we can place limits on the brightness of sources we can study.  For very bright sources, such as exoplanet transits around bright stars, we are limited by the pixel well depth of $10^7$ electrons; if the well depth is achieved in a one second exposure, we can study sources as bright as $J=3.9$, before saturation on an 18-inch telescope at a signal-to-noise level of over 3000 ($0.4$ mmag precision).  For dimmer sources, such as distant SNe or GRBs, our thresholds are set by statistical sources of noise.  In our best case scenario, we are limited by the sky background; this would require lowering our dark, by cooling the detector to $0^{\circ}C$ (Figure \ref{cooling}).  If the sky background is the dominant source of noise, we expect to be able to resolve sources of $J=16.35$ at $10\sigma$ in a one minute exposure with the InGaAs detector mounted to an 18-inch telescope.  The field of view of the detector on our 18-inch ($f/4.5$) telescope is $16'\times21'$.
 
Current generation, off-the-shelf, InGaAs detectors offer a cost-effective way to study the NIR sky, as they do not need the drastic (and therefore expensive) cooling that HgCdTe detectors require.  The low cost of these detectors would make them useful for compound focal planes or to enable arrays of small telescopes each with single or a few detectors.  It would, therefore, be possible to build up sky coverage for monitoring multiple bright sources or for conducting wide-field, sky-limited (but relatively shallow) surveys in the NIR.  Both of these science cases would benefit from the large detector pixels of the device we have studied.  The large well-depth allows for monitoring of very bright sources, while the large pixels (i.e. on the sky) allow us to potentially reach the sky-background limit with modest cooling.

\section{Acknowledgements}
This publication makes use of data products from the Two Micron All Sky Survey, which is a joint project of the University of Massachusetts and the Infrared Processing and Analysis Center/California Institute of Technology, funded by the National Aeronautics and Space Administration and the National Science Foundation.

Some of the data presented in this paper were obtained from the Mikulski Archive for Space Telescopes (MAST). STScI is operated by the Association of Universities for Research in Astronomy, Inc., under NASA contract NAS5-26555. Support for MAST for non-HST data is provided by the NASA Office of Space Science via grant NNX09AF08G and by other grants and contracts.

\end{document}